\documentstyle[epsf,eqsecnum,floats,preprint,aps,epsfig]{revtex}

\input epsf
\tighten
\def\mpl{M_{\rm Pl}}

\begin{document}
\newcommand{\bm}[1]{\mbox{\boldmath{$#1$}}}
\newcommand{\be}{\begin{equation}}
\newcommand{\ee}{\end{equation}}
\newcommand{\bea}{\begin{eqnarray}}
\newcommand{\eea}{\end{eqnarray}}
\newcommand{\barr}{\begin{array}}
\newcommand{\earr}{\end{array}}

\rightline{}
\rightline{hep-th/0408084}
\rightline{UFIFT-HEP-04-11}
\vskip 1cm

\begin{center}
\ \\
\large{{\bf Multi-Throat Brane Inflation }} 
\ \\
\ \\
\ \\
\normalsize{ Xingang Chen }
\ \\
\ \\
\small{\em Institute for Fundamental Theory \\ Department of Physics,
University of Florida, Gainesville, FL 32611 }

\end{center}

\begin{abstract}
We present a scenario where brane inflation arises more
generically. We start with D3 and anti-D3-branes at the infrared ends
of two different throats. This setup is a natural consequence of the
assumption that in the beginning we have a multi-throat string
compactification with many wandering anti-D3-branes. A long period of
inflation is triggered when D3-branes slowly exit the highly warped
infrared region, under a potential generically arising from the moduli
stabilization. In this scenario, the usual slow-roll conditions are
not required, and a large warping is allowed to incorporate the
Randall-Sundrum model.
\end{abstract}

\setcounter{page}{0}
\thispagestyle{empty}
\maketitle

\eject

\vfill

\baselineskip=18pt

\section{Introduction}
Brane inflation\cite{Dvali:1998pa,Dvali:2001fw} is an interesting idea
to construct the inflationary
models in string theory. Recent studies\cite{Kachru:2003sx,Hsu:2003cy}
have shown that such models can
be realized in warped space. The setups typically have anti-D3-branes
sitting at the infrared (IR) cut-off of the warped space (throat),
while the 
position of the mobile D3-branes, which enter from the ultraviolet
(UV) entrance, plays the role of the scalar
field. Inflation can be achieved when the slow-roll conditions are
satisfied, and ends by brane-anti-brane collision and
annihilation. A common feature for these models is that
various amount of fine-tuning of different parameters is required.
The reason traces back to the usual slow-roll conditions which demand
a very flat potential. Given the rich stringy phenomena that have been
discovered recently, it would certainly be interesting to look for
more generic situations where the inflation can happen. We will
discuss such a model in this paper.

We first notice that the slow rolling of the transverse brane motion
can occur in a warped space as well without a flat potential,
essentially because the 
coordinate speed of light along the transverse direction decreases due 
to the warping. Silverstein, Tong\cite{Silverstein:2003hf} and
Alishahiha\cite{Alishahiha:2004eh} have recently made use
of this fact and studied a scalar coupled inflationary cosmology with
a non-trivial 
kinetic term, where the rolling scalar corresponds to the position of
a D3-brane moving toward the IR end of a throat. However, because of
the
back-reaction of the relativistic
D3-brane\cite{Silverstein:2003hf,Chen:2004hu}, the D3-brane quickly
enters a non-comoving phase\cite{Chen:2004hu} and most of the IR
portion of the
warped space becomes irrelevant for our purpose.

Different from the previous setups, we start
our mobile
D3-branes from the IR end of another throat. We will explain how
this situation can arise rather naturally from a multi-throat string
compactification with many
wandering anti-D3-branes. 
``Slow-roll'' happens when the D3-branes exit the exponentially warped
IR
end under a generic quadratic potential with a right sign. The
detailed form of
the potential is not very important. We will also find
that this scenario allows an exponentially large warping, thus the
Randall-Sundrum model\cite{Randall:1999ee}
can be incorporated. We will discuss
some subtleties involved in the reheating process.

\section{Throat construction and brane-flux annihilation}
In this section, we briefly introduce some background material on
warped space in string compactification which
will be useful to set up our scenario.

The simplest way to create a warped space in string compactification
is to place a stack of $N$ D3-branes transverse to the compactified
dimensions\cite{Verlinde:1999fy}. A probe brane moving in the
near-horizon region of these
D3-branes will feel an AdS background. If all the D3-branes are
coincident, the warp factor goes all the way to zero in the IR
direction. There is no
mechanism to stabilize a large but finite hierarchy in this setup.

As shown in\cite{Klebanov:2000hb,Giddings:2001yu}, this can be
achieved by replacing the D3-branes with
three-form fluxes near a conifold singularity on a Calabi-Yau
manifold. If there are $M$ units of RR fluxes and $K$ units of NSNS
fluxes
on two dual three-cycles respectively, we will have a warped space
with a
smooth end and minimum warp factor
\bea
h_{\rm min} \sim \exp(-2\pi K/3M g_s) ~,
\label{hmin}
\eea
where $g_s$ is the string coupling. Therefore a large hierarchy can be
obtained from a moderate number of fluxes. Away from the IR tip
region, the resulting geometry is approximately AdS and the
characteristic length
scale $R$ is related to the flux number by $R^4=\frac{27}{4} \pi g_s N
\alpha'^2$ with $N=MK$.

Introducing anti-D3-branes at the IR end of the throat breaks the
supersymmetry\cite{Kachru:2002gs}. This has been used by
KKLT\cite{Kachru:2003aw} to lift the AdS vacuum
and construct the dS space. If the warped space is generated by
D3-branes, we expect some of them to be annihilated by the
anti-D3-branes. But in the case of the three-form fluxes, there is no
obvious way to annihilate the anti-D3-branes. A very
interesting process of brane-flux annihilation is described
in\cite{Kachru:2002gs,DeWolfe:2004qx}. What happens is that, starting
with $p~(p<M)$ number of
anti-D3-branes in the IR tip,  they tend to cluster and
then puff up into a fuzzy NS5-brane in the presence of the
fluxes. This NS5-brane will unwind an internal three-cycle,
annihilating one unit of NSNS flux and creating $M-p$ D3-branes to
conserve the total D3-charge. The unwinding process can happen either
through quantum tunneling if $p\ll M$, or through classical rolling
otherwise.

\section{Brane dynamics in a throat}
A detailed description of the brane dynamics in warped space
will be very important to our scenario.
In this paper, we are interested in the dynamics of a D3-brane exiting
from the IR end of a
throat in a background four-dimensional dS space. We
approximate the warped space
as a simple AdS space
\bea
ds^2 = h^2(r)\left( -dt^2 + a^2(t) d{\bf x}^2 \right) + h^{-2}(r)dr^2
~,
~~~ h(r)= r/R
\label{AdS5}
\eea
with a four-form potential $C_{0123}= -h^4 a^3$.
The $a(t)$ is the scale factor of the dS space.
The angular motion in the extra dimensions is neglected.
We will start with a static D3-brane in
the IR region. For
D3-brane, the gravitational and four-form potentials
cancel each other. But in string compactification, a potential
will 
arise from the volume stabilization\cite{Kachru:2003sx}. We take such
a potential to be
\bea
V(r)= -\frac{1}{2} H^2 r^2 
\label{PotentialVm}
\eea
in unit of brane tension $T_3$, where $H$ is the Hubble constant of
the four-dimensional inflationary space. The absolute magnitude of this
potential is of the 
same order as that arises from the conformal
coupling\cite{Kachru:2003sx}. In the brane
inflation considered in \cite{Kachru:2003sx,Hsu:2003cy}, there are
different sources of such mass
terms. The slow-roll conditions require cancelations among them to a
certain precision. Here we will just take the generic magnitude, but
choose the sign to make sure that the D3-brane will exit the throat.

The homogeneous probe D3-brane obeys the following DBI action
\bea
S=T_3\int d^4x \left[ -a^3(t) h^2 \sqrt{h^4-\dot r^2} +
a^3(t)\left(h^4 - V(r) \right) \right] ~.
\label{DBIS}
\eea
The second term in
(\ref{DBIS}) includes both the Chern-Simons potential from the RR
five-form field and the potential (\ref{PotentialVm}). 
Under the acceleration of the potential
(\ref{PotentialVm}), for $t \ll -H^{-1}$, the equation of motion gives
the behavior of the D3-brane as
\bea
r = -\frac{R^2}{t} + \frac{9R^2}{2H^2 t^3} +
\cdots ~,
\label{rAsymp}
\eea
where the leading order is that of light and the second order is
determined by the
potential. Note here that we have chosen the time $t$ to run from
$-\infty$. In terms of $r$, the valid region for (\ref{rAsymp})
becomes 
\bea
r \ll R^2 H ~.
\label{rRange1}
\eea

The probe back-reaction can be characterized
by the Lorentz contraction factor $\gamma
= h^2/\sqrt{h^4-\dot r^2}$ of the relativistic
D3-brane\cite{Silverstein:2003hf,Chen:2004hu}. It has to be
much smaller than the
strength of the background. This requires $\gamma \ll N$, because the
background can be thought of as being created by $N$
D3-branes. Using the behavior (\ref{rAsymp}), this requires
\bea
r \gg \frac{1}{3N} R^2 H ~.
\label{rRange2}
\eea
We note that, while the lower bound (\ref{rRange2}) for $r$ is
the consistency requirement to use the probe dynamics, the upper bound
(\ref{rRange1}) is not a constraint -- it gives a region where we can
simply treat the brane as being highly relativistic, outside of which
the D3-brane can only move slower than the speed of light.

\section{Our scenario}
Now we are ready to describe our scenario. We assume that in the
beginning
we have a multi-throat configuration in the extra dimensions,
resulting from a string compactification with three-form
fluxes. This can be regarded as a generalization of the
KS-GKP\cite{Klebanov:2000hb,Giddings:2001yu}
model, but an explicit construction is beyond the scope of this
paper. We start with many wandering anti-D3-branes in the bulk. These
anti-D3-branes will be attracted to and settle down in the IR ends of
different throats. Depending on the number of fluxes and
anti-D3-branes associated with
each throat, the anti-D3-branes will have different lifetime. For
example, some of them may decay earlier as a result of brane-flux
annihilation; the corresponding throat may become more shallow or even
disappear; the resulting D3-branes may stay or come out of the throat
and enter the other throats. During these processes, the usual
slow-roll inflation may also happen under some special conditions.

Here we are most interested in the last step. Suppose there exits a
throat with relatively large warping. After the anti-D3-branes
annihilate with some number of NSNS fluxes either quantum mechanically
or classically, the resulting D3-branes take the longest time to come
out. This can happen either because the decay process is the slowest,
or the D3-branes move most slowly due to the large warping as
described in the last section. 
We will then concentrate on the dynamics of these D3-branes
alone since all other processes have already happened. We will call
this throat the brane-throat (B-throat).
As mentioned, the gravitational and four-form potentials cancel each
other for D3-branes. So for them to get out of the throat, a potential
with a right sign such as (\ref{PotentialVm}) is needed.

The inflationary
energy is provided by the anti-D3-branes in some other throats, where
they have relatively longer lifetime -- in fact, among them only
that with the biggest warp factor provides the dominant inflationary
energy, because the energy density of the others are red-shifted.
The long period of inflation is triggered by the slow motion of the
D3-branes in the deep B-throat.

After the D3-branes finally come out of the B-throat, they will
then be attracted to different throats under similar potentials as
(\ref{PotentialVm}) but with an opposite sign. They will annihilate
some of the anti-D3-branes and end the inflation. Among all these
throats, there are two of them which are especially
interesting.
One provides the dominant inflationary energy as mentioned above,
which we will call
the anti-brane-throat (A-throat); and another is where
the standard model eventually lives, which we will call the
standard-model-throat (S-throat). In the simplest case, the A-throat
and S-throat can be the same throat.

All the processes are happening under the assumption that the shape
and size moduli of the compactified dimensions are fixed, and 
that in the end of the inflation our universe is in
a meta-stable dS space with a small cosmological
constant. This is in the sense of the KKLT\cite{Kachru:2003aw} and
KKLMMT\cite{Kachru:2003sx} models.
We will use the subscripts $B$, $A$ and $S$ to label the quantities
related to
B-throat, A-throat and S-throat,
respectively.

\section{Inflationary e-folds and density perturbation}
We now calculate the resulting inflationary e-folds. The
anti-D3-branes in the A-throat provide the
inflationary energy density\cite{Kachru:2003sx} $V_A = 2 T_A h_A^4$,
where $T_A=n_A T_3$ is the total brane tension of $n_A$
anti-D3-branes and $h_A$ is the warp factor. A factor of two comes in
due to the summation of the gravitational and four-form potentials. (A
possible shift from the potential
(\ref{PotentialVm}) may be included in the $n_A$ which is then no
longer an integer.) Here we consider one D3-brane in the B-throat.
We denote the warp factor at 
the position of
the D3-brane as $h_B=r_B/R_B$. Starting from $r_B$, all of
the inflationary e-folds $N_e$ are obtained within a range of order
$N_e r_B$. One
can check that the change of the potential (\ref{PotentialVm}) within
this range is negligible. So the inflationary energy is nearly a
constant during the inflation. We can then approximate $N_e$ as
\bea
N_e = \int H dt \approx H R_B h_B^{-1} \approx \sqrt{\frac{2}{3}}
\frac{\sqrt{T_A} R_B}{\mpl} h_A^2 h_B^{-1} ~.
\label{Ne}
\eea

Using (\ref{Ne}) we can rewrite the conditions (\ref{rRange1}) and
(\ref{rRange2}) as (with $N$ specified as $N_B$)
\bea
\frac{N_e}{3N_B} r_B \ll r \ll N_e r_B ~.
\label{rRange}
\eea
The upper bound means that, in most of the inflationary period, this
D3-brane is relativistic. This justifies our approximation in
(\ref{Ne}) where we estimate $\Delta t \approx R_B h_B^{-1}$ using
(\ref{rAsymp}). This can be readily understood as follows. Since our
potential (\ref{PotentialVm}) does not satisfy the usual slow-roll
conditions, the slow-roll behavior has to come from the causality
constraint. This is the reason that the D3-brane travels nearly at the
speed
of light. The lower bound gives the maximum number of the e-foldings,
up to $N_B$,
that can be achieved in this model without triggering the
back-reaction. So $N_e$ denotes the latest
e-folds.\footnote{Since a signal takes a time of $N_e H^{-1}$ to
reach $r_B$ from the bulk,
it is also necessary to start the inflation earlier in order to
assume that the Hubble constant is independent of the extra
dimensions.}

In terms of the fundamental parameters in the expression (\ref{Ne}), a
long period of inflation can be obtained easily. Using
$T_3=(2\pi)^{-3} g_s^{-1} \alpha'^{-2}$ and $R_B^4 =\frac{27\pi}{4}
g_s N_B \alpha'^2$, Eq.~(\ref{Ne}) becomes
\bea
N_e \approx 0.1 \sqrt{n_A} ~ g_s^{-1/4} N_B^{1/4} ~
\frac{\alpha'^{-1/2}}{\mpl} ~ h_A^2 h_B^{-1} ~.
\eea
For example, if $0.1 \sqrt{n_A}~ g_s^{-1/4} N_B^{1/4} \alpha'^{-1/2}/
\mpl \sim 1$, then we need $h_B \sim
h_A^2/N_e$.
In terms of the flux number
in (\ref{hmin}), this is easy to satisfy.\footnote{From here we also
see that the energy density of
the moving D3-branes, which is $T_B h_B^4
(\gamma_B-1)
\approx \frac{1}{3} N_e T_B h_B^4$, is negligible comparing to the
inflationary energy density $V_A$, as assumed in (\ref{DBIS}).}

In a dS space, the amplitude of the scalar quantum fluctuations being
stretched out of the horizon is $H/2\pi$. Translated to the
transverse brane coordinate $r_B$,  this gives the perturbation
$\delta r_B = H/(2\pi \sqrt{T_3}) \approx r_B N_e/\tilde N$, 
where $\tilde N \equiv (27 N_B/8)^{1/2}$.
(Note that the Lorentz contraction $\gamma_B^{-1}$ in the transverse
direction is cancelled by a dilation factor $\gamma_B$ which gives a
Hubble constant $H\gamma_B$ for a moving observer on the D3-brane.)
These ripples on the D3-brane
cause a spatially dependent delay $\delta t$ for the ending of the
inflation to an unperturbed observer. Using (\ref{rAsymp}) and
(\ref{Ne}), this
gives the density perturbation\cite{Peebles:1994xt}
\bea
\delta = \frac{2}{5} H_r \delta t 
\approx \frac{2}{5} H_r R^2 \left(
\frac{1}{r_B} - \frac{1}{r_B+\delta r_B} \right) 
\approx \frac{2 H_r}{5 H}
\frac{N_e^2}{\tilde N +N_e} ~,
\label{DP}
\eea
where $H_r$ is the Hubble constant during the reheating and is
proportional to the squared warp factor of the S-throat
$h_S^2$.

In this
section we assume that matter created during the reheating comes
from the energy density released from the brane-anti-brane
annihilation, and the conversion is efficient. 
In the simplest case where the A-throat and
S-throat are identical, this implies $H_r \approx
H$. For a normal value of $\tilde N$, this gives a too big density
perturbation. Therefore we consider the case where they are
different. In our scenario the reheating on the standard-model
anti-D3-branes is not a problem even if the S-throat is different from
the A-throat, since the inflation is not caused by
D3-branes moving near the UV entrance of the A-throat as in
KKLMMT. In our case, there can be many D3-branes coming out of the
B-throat. They will then
scatter into different throats, including the
S-throat, colliding and annihilating anti-D3-branes there
as
a reheating mechanism. (But to ensure that the subsequent cosmological
evolution is dominated by the S-throat, the anti-D3-branes
in the A-throat have to be all annihilated, and the possible extra
D3-branes will have to come out and enter other throats, including the
S-throat, which have $h \lesssim h_S$.)

For example, the experiments indicate that the density perturbation in
(\ref{DP}) is of order $10^{-5}$. If we take $\tilde N \sim 100$ and
$N_e \sim 60$, this gives $H_r/H \sim 10^{-6}$. That is, ignoring the
factors $n_S$ and $n_A$, the
S-throat has a warp factor roughly seven e-folds smaller than
the A-throat. Since only the relative ratio $h_S^2/h_A^2$ appears in
(\ref{DP}), extremely small numbers are
allowed for $h_S$ and $h_A$. Therefore our model can incorporate the
Randall-Sundrum model\cite{Randall:1999ee} to solve the hierarchy
problem.

As we can see, certain details discussed above,
including whether the S-throat can be identified with the
A-throat,
depends on the magnitude of the density perturbation in (\ref{DP}). As
we will discuss in the next section, there are some subtleties
regarding the reheating process in the deep throat. So further
investigation is needed.\footnote{A more detailed
study\cite{Chen:2005ad} shows that
such subtleties will modify the dependence of Eq.~(\ref{DP}) on warp
factors.}

We now turn to some quantities which are less dependent of the overall
magnitude of the density perturbation -- the spectral index $n_s$ and
the
running of this index. Since the number of the latest e-folds $N_e$ is
related to the wave-number $k$ by
$d\ln k = -dN_e$, we get
\bea
n_s -1 \equiv 2 \frac{d\ln \delta}{d\ln k} 
\approx -\frac{4}{N_e} +\frac{2}{\tilde N +N_e} ~,
\label{ns} \\
\frac{d n_s}{d\ln k} 
\approx -\frac{4}{N_e^2} +\frac{2}{(\tilde N+N_e)^2}~.
\label{nsRun}
\eea
These give a red-tilted scale invariant spectrum with negative
running. Modifications of (\ref{DP}), (\ref{ns}) and (\ref{nsRun}) may
come from the modification of the approximate geometry in
(\ref{AdS5}), especially in the tip region of the
throat\cite{Herzog:2002ih}. But perhaps the
most interesting possibility comes from the following consideration.

Consider the case $\gamma_B^2 H\delta t \gtrsim 1$, which happens for
$N_e \gtrsim
\sqrt{3} \tilde N^{1/4}$. In this case, for a moving observer on
D3-branes the transverse brane
fluctuations ($\gamma_B \delta r_B \approx \gamma_B \delta t h_B^2$),
generated
quantum mechanically from the vacuum in a Hubble time ($\gamma_B^{-1}
H^{-1}$), 
travels faster than the speed of
light ($h_B^2$) in the transverse direction. This is impossible. 
The reason is
that the field theory
description on the D3-branes has broken down. It is unusual because
the Hubble energy scale is now approximately greater than the
red-shifted string
scale, i.e.~$\gamma_B H \gtrsim \sqrt{2\pi} T_B^{1/4} h_B$, due to the
large warping in the B-throat. The amplitudes of the stringy
quantum fluctuations should be bound by the causality, but details are
not well studied so far. This is an interesting case where the
stringy effects are directly responsible for seeding the structure
formation.

\section{Discussion}
Reheating for brane inflation in warped space is usually
caused by brane-anti-brane
annihilation in inflationary energy scale, as we followed in the last
section. (In some sense, Eq.~(\ref{DP}) is already different since
$H_r$ and $H$ can be very different.) In this
section, we 
discuss some subtle effects\cite{Chen:2004hu} which are different from
the conventional field theory expectation due to fast-rolling
D3-branes in the S-throat. They
affect the overall magnitude of the density perturbation, and the
relation between $N_e$ and $k$.

There are two different processes -- collision and annihilation --
that create open strings on the residue anti-D3-branes. 
For annihilation\cite{Sen:2002nu} this is only possible from loop
diagrams connecting
the decaying branes and anti-D3-branes\cite{Chen:2003xq}. The resulted
energy density cannot exceed
$2n_S T_3 h_S^4$, where $n_S$ is the number of anti-D3-branes
annihilated in the S-throat and $h_S$ is the corresponding
warp factor. In addition, they have to compete with the
disk\cite{Lambert:2003zr} and loop
diagrams which create closed strings to the bulk.

We next look at the collision. As studied in
\cite{Silverstein:2003hf,Chen:2004hu}, when the D3-branes
enter from the UV entrance and move toward the IR end, their kinetic
energy density may approach $N_S$ 
times the brane tension and enter a non-comoving
phase\cite{Chen:2004hu}. Then their
final energy density cannot grow too much greater. This is
relatively independent of
the total warping or the initial velocity, and is determined by the
strength of background geometry characterized by $N_S$. The
collision will transfer an energy density of order $N_S T_3
h_S^4$, which is dominant over that from the annihilation since $N_S$
is a big number. This may be
a good news -- the collision can be a much more efficient
process to create open strings especially if the corresponding
oscillating scalar
has bigger coupling to matter than gravity.

Furthermore, at the beginning of the reheating, the effective warp
factor at the IR end is much greater than its original value, due to
the
back-reaction. After things cool down, it restores to $h_S$. This has
the rescaling effect\cite{Chen:2004hu} (to a Poincare observer) which
stretches the wavelength of the density
perturbation by a ratio that can be as big as several e-folds below
the total warping. On the branes, if we stay in a frame in which the
time and length do not rescale, the effective Planck mass increases
from an exponentially low value to the present value, as the IR warp
factor restores. It is certainly interesting to see if any dramatic
effects can be caused to the density perturbation.

\acknowledgments 
This work was supported in part by the Department of
Energy under Grant No.~DE-FG02-97ER-41029.

\end{document}